# Two-dimensional time- and space-resolved diagnostic method for integrated implosion process

SHIJIAN LI,[1] QIANGQIANG WANG,[2] XURI YAO,[1] ZHURONG CAO,[2,5] JIE LIU,[3] AND QING ZHAO[1,4,6]

[1] *Center for Quantum Technology Research and Key Laboratory of Advanced Optoelectronic Quantum Architecture and Measurements (MOE), School of Physics, Beijing Institute of Technology, Beijing 100081, China*
[2] *Laser Fusion Research Center, China Academy of Engineering Physics, Mianyang, Sichuan 621900, China*
[3] *Graduate School of China Academy of Engineering Physics, Beijing 100193, China*
[4] *Beijing Academy of Quantum Information Sciences, Beijing 100193, China*
[5] *cao33jin@aliyun.com*
[6] *qzhaoyuping@bit.edu.cn*

**Abstract:** To precisely measure and evaluate X-ray generation and evolution in a hohlraum during an implosion process, we present a two-dimensional (2D) time- and space-resolved diagnostic method by combining a compressed ultrafast photography (CUP) system and a simplified version of space-resolving flux detector (SSRFD). Numerical experiment results showed that the reconstruction quality of the conventional CUP significantly improved owing to the addition of the external SSRFD, especially when a coded mask with a large pixel size was used in the CUP. Further, the performance of the CUP cooperation with the SSRFD was better than that of adding an external charge-coupled device or streak camera. Compared with existing ultrafast imaging techniques in laser fusion, the proposed method has a prominent advantage of measuring the 2D evolution of implosion by combining high temporal resolution of streak camera and high spatial resolution of SSRFD; moreover, it can provide guidance for designing diagnostic experiments in laser fusion research.

## 1. Introduction

In the study of inertial confinement fusion (ICF), the precise measurement of X-ray generation and evolution plays a crucial role in controlling physical experiment parameters and verifying simulation programs. With the development of X-ray streak and X-ray framing cameras [1–4] specifically designed for ICF measurements, these devices can achieve ultrahigh temporal resolution (TR) and spatial resolution (SR) of X-ray diagnosis [5–8]. However, the high TR of X-ray streak cameras is achieved at the expense of imaging dimension. The streak cameras used in laser fusion limit their imaging to only one spatial dimension. The X-ray framing cameras can simultaneously provide time- and space-resolving X-ray diagnostics; however, owing to its limited frame number, it cannot achieve ultrahigh TR in the implosion process. In 2015, a new two-dimensional (2D) space-resolving flux detection technique was developed to measure X-ray flux inside a hohlraum using a time- and space-resolving flux detector (SRFD) [9] and successfully conducted in the China laser facility [10].

In the last few decades, high-speed imaging has made significant progress, especially with the invention of high-performance image sensors based on charge-coupled device (CCD) and complementary metal-oxide semiconductor, high-speed imaging speed has been able to achieve $10^7$ frames per second (fps) [11]. Such frame rates have been unable to meet the increasing demand for ultrafast imaging. Recently, the pump-probe technique has become the dominant method for capturing transient events [12]; however, it requires that the captured events are repeatable. Capturing nonrepeating time evolution events at ultrahigh speed is essential, especially in ICF. Compressed ultrafast photography (CUP) [13], as a receiving-only ultrahigh-speed imaging technology, was proposed in 2014. Such a method realized 2D computational ultrafast imaging on the basis of streak camera and compressive sensing. Related work has been proposed and effective results have been obtained, continuously improving the CUP—from monochromatic to multispectral imaging [14–16] and from visible to ultraviolet light [17]; the reported highest frame rate has been able to achieve $7 \times 10^{13}$ fps [18]. While achieving higher TR, SR is insufficient owing to the existence of encoding and decoding steps.

In this paper, we present a 2D time- and space-resolved diagnostic method for the integrated implosion process by combining a CUP system and a simplified version of SRFD (SSRFD). Fig. 1(a) shows the schematic diagram of the proposed method. Measurement via the CUP system is imaged through a pinhole on the left side of the hohlraum, whereas the SSRFD measurement is obtained through a pinhole on the right side of the hohlraum. Finally, the dynamic scenes of the integrated implosion process are reconstructed from these measurements by adopting a compressive sensing algorithm. To the best of our knowledge, no 2D time- and space-resolved diagnostic method with high resolution in the entire implosion process exists. Numerical experiment results show that the proposed method achieves better reconstruction quality than the conventional CUP. Compared with the CUP, the high TR of the streak camera was retained, and insufficient SR of CUP significantly improved by adding the SSRFD.

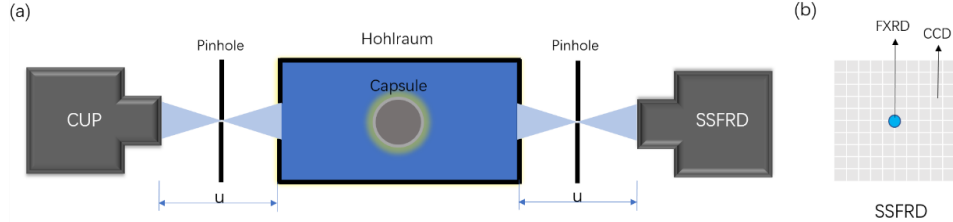

Fig. 1. Schematic of the proposed method: (a) complete schematic of the system, (u is the distance between the laser entrance hole and imaging plane); (b) detailed illustration of the SSFRD.

## 2. Method

### 2.1 CUP

In the conventional CUP system, dynamic scenes $I(x,y,t)$ are imaged via an objective lens and a 4f imaging system, spatially modulated by a digital micromirror device (DMD), and then measured using a streak camera. The measured image $E(m,n)$ using the conventional CUP system can be mathematically expressed as follows [13]:

$$E(m,n) = TSCI(x,y,t) \qquad (1)$$

where $C$ is the spatial-encoding operator, $S$ is the temporal-shearing operator, and $T$ is the spatiotemporal-integrating operator. Given the information on the coded mask, the inverse problem of Eq. (1) is solved to reconstruct the dynamic scenes from the CUP measurement. This process can be formulated as follows:

$$\arg\min_{I}\{\tfrac{1}{2}(E-TSCI)+\lambda TV(I)\} \qquad (2)$$

where TV () denotes the total variation (TV) regularizer, and $\lambda$ is the regularization parameter. Eq. (2) can be solved by adopting a compressive sensing algorithm.

### 2.2 SSRFD

The concept of SRFD was first proposed by Kuan Ren et al. in 2015 [9]. The SSRFD comprises an X-ray CCD at the periphery and a flat-response X-ray detector (FXRD) [19] at the center (Fig. 1(b)). The size of the SSRFD is the same as that of CCD in the streak camera. The peripheral CCD provides the time-integrated image of dynamic scenes with high SR, whereas the internal FXRD provides the X-ray flux information with TR. The diameter of internal FXRD is one-tenth that of the SSRFD. The internal FXRD can be customized and placed in an area where the dynamic scenes change the most with time. In this study, the internal FXRD was placed at the center of the SSRFD.

*2.3 Combination of the CUP and SSRFD*

To reconstruct a 2D dynamic scene with high TR and SR, the measurement system is composed of a CUP and an SSRFD (Fig. 1(a)). The data $E$ using the proposed method can be mathematically expressed as follows:

$$E = \begin{cases} E_{CUP} = TSCI(x,y,t) \\ E_{SSRFD} = \begin{cases} E(m,n) = TI(x,y,t), m,n \in CCD \\ \sum_{m,n \in FXRD} \int_{(i-1) \cdot t_0}^{i \cdot t_0} I(x,y,t) dt \end{cases} \end{cases} \quad (3)$$

where $C$ is the spatial-encoding operator, $S$ is the temporal-shearing operator, and $T$ is the spatiotemporal-integrating operator; $i$ is the *i-th* time interval, and $t_0$ is the TR of the SSRFD. The measured data $E$ contain two parts: the measurement of the CUP and SSRFD. Similar to the conventional CUP, the dynamic scenes can be reconstructed from the measurement by adopting a compressive sensing algorithm. In this study, a generalized alternating projection algorithm with a TV regularizer [20] is adopted to reconstruct the dynamic scenes, and a BM3D [21] denoiser is used after every 50 iterations to obtain better reconstruction results.

## 3. Numerical experiments setup

A $100 \times 100$ image with two Gaussian distributions was used as a base image, corresponding to the physical size of 1 mm × 1 mm, which means the CCD's pixel size is 10 μm × 10 μm. We consider a dynamic scene set with 80 frames and a total time of 2 ns with a TR of 25 ps. Similar to the result in Ref. [22], the maximum intensity of each frame with respect to time is shown in Fig. 2(a). The variances of the Gaussian distribution along the x and y-axis in each frame are slightly different so that the difference in the radiation field distribution among frames is not only the intensity but also the outline. (Figs. 2(b–d)). A 1% Gaussian white noise was added to the measurement.

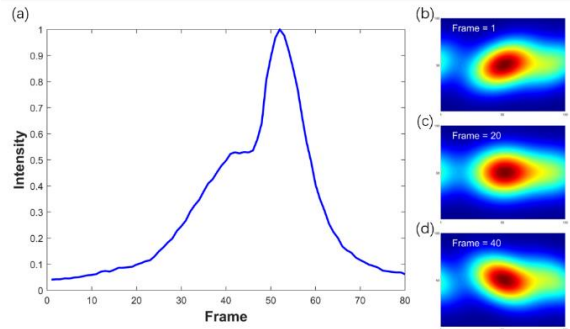

Fig. 2. Numerical experiments setup of intensity: (a) maximum intensity of each frame; (b–d) different intensity distributions of three frames.

*3.1 Validation test*

In the conventional CUP, the coded mask is usually a DMD displayed at a pseudorandom binary pattern. In laser fusion, DMD is unsuitable as a coded mask for X-ray. We consider using a transmission mask with a large pixel size because of manufacturing difficulties. In this test, we use a mask of 10 times CCD unit pixel, that is, a 10 × 10 mask with a pixel size of 0.1 mm × 0.1 mm. The TR of SSRFD is the same as that of SRFD used in Ref. [9], which is 100 ps.

*3.2 Comparison tests*

To illustrate the advantages of our method, we set up four groups of numerical experiments, the first was the conventional CUP (marked as CUP), the second was two streak cameras using complementary masks (marked as Two_SC), the third was CUP with an external CCD camera placed at the right of a hohlraum (marked as SC_CCD), and the last was the proposed method with different TR for the SSRFD. The contribution to reconstruction quality from the pixel size of coded mask was also investigated in three cases: 10 × 10, 20 × 20, and 100 × 100 coded masks (pixel sizes of 0.1 mm × 0.1 mm, 50 μm × 50 μm, and 10 μm × 10 μm, respectively).

## 4. Results and discussion

*4.1 Validation test*

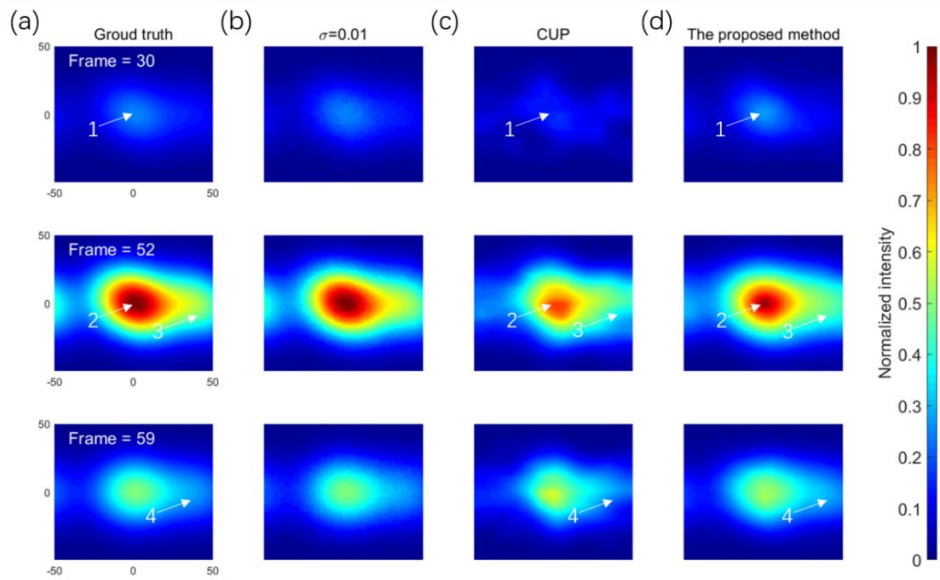

Fig. 3. Comparison of three frames: (a) simulated ground truth and (b) ground truth with Gaussian white noise $\sigma = 0.01$. (c, d) reconstructed frames of the conventional CUP system and the proposed method, respectively.

Fig. 3 shows three frames of the ground truth and the reconstructed intensity profile; the general shape and intensity of distributions are all maintained using the proposed method. Focusing on the regions marked in the figure, the intensities of the proposed method were closer to the ground truth. The reconstructed intensity distribution of the conventional CUP distorted owing to noise and the large pixel size of the mask, but the proposed method does not. The proposed method achieves better smoothness. To quantitatively assess the quality of recovery, the peak signal-to-noise ratio (PSNR) is defined. The PSNR between the source x and the restored y is defined as follows:

$$PSNR(x, y) = 10\log_{10}(\frac{peakval^2}{MSE(x, y)}) \quad (4)$$

where $MSE(x, y)$ is the mean square error between x and y. the $peakval$ is the maximum value of the image data type. The larger the value of PSNR, the higher is the reconstruction quality. Moreover, to evaluate the similarity of the reconstructed dynamic scene, the structural similarity (SSIM) is defined as follows [23]:

$$SSIM(x, y) = \frac{(2\mu_x\mu_x + C_1)(2\sigma_{xy} + C_2)}{(\mu_x^2 + \mu_y^2 + C_1)(\sigma_x^2 + \sigma_y^2 + C_2)} \quad (5)$$

where $\mu_x$ is the mean of x, $\sigma_x$ is the variance of x, $\sigma_{xy}$ is the covariance of x and y, and $C_1$, $C_2$ are constants. The value range of SSIM is [0, 1]. The larger the value of SSIM, the higher the SSIM.

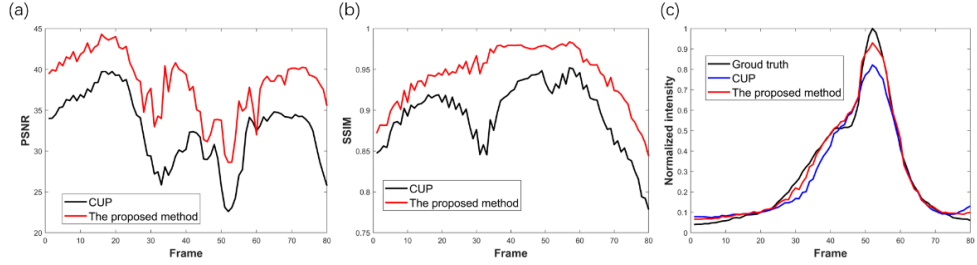

Fig. 4. comparisons of the conventional CUP with the proposed method for different frames: (a) PSNRs of different frames; (b) SSIMs of different frames. (c) Normalized total intensity of different frames.

The PSNR between ground truth and reconstructed intensity distribution was calculated for the conventional CUP and the proposed method as 30.3833 and 36.4277, respectively. The PSNRs of all frames were calculated to evaluate the reconstruction difference during the implosion (Fig. 4(a)). The proposed method was superior to the conventional CUP at each point. Further, the SSIM of the proposed method was 0.9329, whereas that of the conventional CUP was 0.8797. Fig. 4(b) shows the SSIMs of all frames, which has the same trend as Fig. 4(a). That is, the proposed method was superior to the conventional CUP.

In indirectly driven laser fusion, the measurement of the radiation flux crucial in the diagnostic process. To compare the reconstruction quality of the two methods across frames, we plotted the normalized total intensity against the frame index (Fig. 4(c)). The reconstruction of the proposed method got closer to the ground truth than the conventional CUP, demonstrating a better reconstruction performance in the time domain.

### 4.2 Comparison tests

To quantitatively evaluate the reconstruction quality of the four methods, we calculate PSNR and SSIM in different cases, and the results are shown in Fig. 5. The PSNRs of the first three methods were 30.3833, 32.4503, and 32.3161, respectively. From Fig. 5(a), when the TR of SSRFD exceeded 500 ps, the PSNR of the proposed method was higher than that of the first three methods. The SSIMs of the first three methods were 0.8797, 0.9213, and 0.9109, respectively. From Fig. 5(b), when the TR of SSRFD exceeded 250 ps, the SSIM of the proposed method was higher than that of the first three methods. The above result implies that when the TR of SSRFD was more than 250 ps, the proposed method was better than the other three methods in terms of SSIM and PSNR. This requirement for the TR of SSRFD can be easily achieved. For example, the TR of SRFD used in the diagnostic experiment of Ref. [9] was 100 ps.

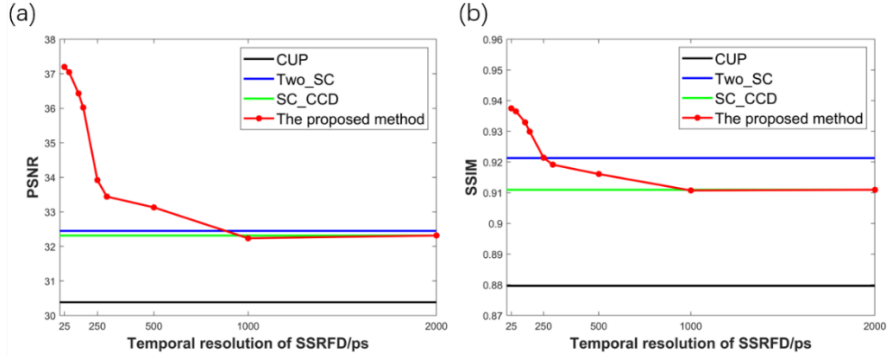

Fig. 5. Comparison of different SSRFD settings: (a) PSNRs of the proposed method under different SSRFD settings; (b) SSIMs of the proposed method under different SSRFD settings.

The mask size used in the above experimental results was 10 times the pixel size of CCD. To evaluate the effect of mask unit size on reconstruction quality, we tested two other mask sizes: one and five times the pixel size of CCD. We calculated the SSIMs and PSNRs of the four methods in different mask settings. The results of our method also showed the performance of the SSRFD with different TRs.

Table 1. PSNRs and SSIMs of the reconstructed dynamic scenes using different methods

|  | Mask size | CUP | Two_SC | SC_CCD | The proposed method ||||| 
|---|---|---|---|---|---|---|---|---|---|
|  |  |  |  |  | Temporal resolution of SSRFD /ps |||||
|  |  |  |  |  | 25 | 50 | 100 | 200 | 250 |
| PSNR | 100×100 | 0.9340 | **0.9516** | 0.9404 | 0.9418 | 0.9416 | 0.9416 | 0.9414 | 0.9333 |
|  | 20×20 | 0.9191 | 0.9345 | 0.9324 | **0.9382** | **0.9377** | **0.9364** | **0.9357** | 0.9333 |
|  | 10×10 | 0.8797 | 0.9213 | 0.9109 | **0.9375** | **0.9365** | **0.9329** | **0.9299** | **0.9214** |
| SSIM | 100×100 | 37.4872 | 40.5037 | 39.7525 | **40.5661** | **40.5448** | **40.5094** | 40.4581 | 36.6491 |
|  | 20×20 | 33.8120 | 35.6289 | 35.1763 | **38.5043** | **38.4160** | **38.0110** | **37.7399** | 36.6491 |
|  | 10×10 | 30.3833 | 32.4503 | 32.3161 | **37.1991** | **37.0462** | **36.4277** | **36.0213** | **33.9204** |

As shown in Table 1, the PSNRs and SSIMs of the reconstructed dynamic scenes improved with the decrease in the minimum pixel size of the mask. The improvements of the first three methods were larger than the last, implying that the first three methods were more dependent on the pixel size of the mask than the proposed method. With the increase in the pixel size of coded masks, the advantages of the proposed method were more. In fusion experiment diagnosis, the pixel size of coded masks is unlikely to be as fine as that of the conventional CUP system for visible light. When the mask size was small, the proposed method was better than the first three methods in terms of both SSIM and PSNR. Moreover, the reconstructed results improved by elevating the TR of the SSRFD.

## 5. Conclusion

In summary, based on the SSRFD and CUP, a 2D space- and time-resolved diagnostic method for integrated implosion targets was developed. Compared with the CUP, the high TR of the streak camera was retained, and insufficient SR of CUP was significantly improved by adding an SSRFD. Situations of coded masks with different pixel size were also investigated. Further, the performance of CUP cooperation with the SSFRD is better than that of adding an external

CCD or streak camera. Numerical experiments showed that when the TR of the entire system was limited by a TR of the streak camera, a smaller pixel size of coded masks or an SSFRD with higher TR significantly improved the SR of reconstruction. In future studies, the proposed method can be further combined with pinhole array imaging, which has potential applications in three-dimensional space- and time-resolved diagnosis.

**Funding**

National Natural Science Foundation of China (NSFC) (11675014, 11675157, and 11805180).

**Acknowledgments.**

The authors would like to thank Prof. Molin Ge for the valuable discussions.

**Disclosures.**

The authors declare no conflicts of interest.

**References**

1. F. Ze, R. L. Kauffman, J. D. Kilkenny, J. Wielwald, P. M. Bell, R. Hanks, J. Stewart, D. Dean, J. Bower, and R. Wallace, "A new multichannel soft x-ray framing camera for fusion experiments," Rev. Sci. Instrum. **63**, 5124-5126 (1992).
2. H. Shiraga, M. Nakasuji, M. Heya, and N. Miyanaga, "Two-dimensional sampling-image x-ray streak camera for ultrafast imaging of inertial confinement fusion plasmas," Rev. Sci. Instrum. **70**, 620-623 (1999).
3. P. Gallant, P. Forget, F. Dorchies, Z. Jiang, J. C. Kieffer, P. A. Jaanimagi, J. C. Rebuffie, C. Goulmy, J. F. Pelletier, and M. Sutton, "Characterization of a subpicosecond x-ray streak camera for ultrashort laser-produced plasmas experiments," Rev. Sci. Instrum. **71**, 3627-3633 (2000).
4. M. Heya, S. Fujioka, H. Shiraga, N. Miyanaga, and T. Yamanaka, "Development of wide-field, multi-imaging x-ray streak camera technique with increased image-sampling arrays," Rev. Sci. Instrum. **72**,755-758 (2001).
5. J. R. Kimbrough, P. M. Bell, G. B. Christianson, F. D. Lee, D. H. Kalantar, T. S. Perry, N. R. Sewall, and A. J. Wootton, "National Ignition Facility core x-ray streak camera," Rev. Sci. Instrum. **72**, 748-750 (2001).
6. J. A. Oertel, R. Aragonez, T. Archuleta, C. Barnes, L. Casper, V. Fatherley, T. Heinrichs, R. King, D. Landers, F. Lopez, P. Sanchez, G. Sandoval, L. Schrank, and P. Walsh, "Gated X-ray detector for the National Ignition Facility," Rev. Sci. Instrum. **77**,10E308 (2006).
7. J. R. Kimbrough, P. M. Bell, D. K. Bradley, J. P. Holder, D. K. Kalantar, A. G. MacPhee, and S. Telford, "Standard design for National Ignition Facility X-ray streak and framing cameras," Rev. Sci. Instrum. **81**,10E530 (2010).
8. C. Trosseille, D. Aubert, L. Auger, S. Bazzoli, T. Beck, P. Brunel, M. Burillo, C. Chollet, J. Gazave, S. Jasmin, P. Maruenda, I. Moreau, G. Oudot, J. Raimbourg, G. Soullié, P. Stemmler, and C. Zuber, "Overview of the ARGOS X-ray framing camera for Laser Megajoule". Rev. Sci. Instrum. **85**, 11D620 (2014).
9. K. Ren, S. Liu, L. Hou, H. Du, G. Ren, W. Huo, L. Jing, Y. Zhao, Z. Yang, M. Wei, K. Deng, L. Yao, Z. Li, D. Yang, C. Zhang, J. Yan, G. Yang, S. Li, S. Jiang, Y. Ding, J. Liu, and K. Lan, "Direct measurement of x-ray flux for a pre-specified highly-resolved region in hohlraum," Opt. Express **23**, A1072-A1080 (2015).
10. K. Ren, S. Liu, X. Xie, H. Du, L. Hou, L. Jing, D. Yang, Y. Zhao, J. Yan, Z. Yang, Z. Li, J. Dong, G. Yang, S. Li, Z. Cao, K. Lan, W. Huo, J. Liu, G. Ren, Y. Ding and S. Jiang, "First exploration of radiation temperatures of the laser spot, re-emitting wall and entire hohlraum drive source," Sci. Rep. **9**, 5050 (2019).
11. Y. Kondo, K. Takubo, H. Tominaga, R. Hirose, N. Tokuoka, Y. Kawaguchi, Y. Takaie, A. Ozaki, S. Nakaya, F. Yano and T. Daigen, "Development of 'Hyper Vision HPV-X' high-speed video camera," Shimadzu Review **69**, 285–291(2012).
12. L. Zhu, Y. Chen, J. Liang, Q. Xu, L. Gao, C. Ma, and L. V. Wang, "Space- and intensity-constrained reconstruction for compressed ultrafast photography," Optica **3**, 694-697 (2016).
13. L. Gao, J. Liang, C. Li, and L. V. Wang, "Single-shot compressed ultrafast photography at one hundred billion frames per second," Nature **516**,74–77(2014).
14. J. Liang, L. Zhu, and L. V. Wang, "Single-shot real-time femtosecond imaging of temporal focusing," Light Sci. Appl. **7**, 42 (2018).
15. J. Yao, D. Qi, C. Yang, F. Cao, Y. He, P. Ding, C. Jin, Y. Yao, T. Jia and Z. Sun, "Multichannel-coupled compressed ultrafast photography," J. Opt. **22**,085701 (2020).
16. C. Yang, F. Cao, D. Qi, Y. He, P. Ding, J. Yao, T. Jia, Z. Sun, and S. Zhang, "Hyperspectrally Compressed Ultrafast Photography," Phys. Rev. Lett. **124**, 023902 (2020).
17. Y. Lai, Y. Xue, C.-Y. Côté, X. Liu, A. Laramée, N. Jaouen, F. Légaré, L. Tian and J. Liang, "Single-Shot Ultraviolet Compressed Ultrafast Photography," Laser & Photonics Rev. **14**, 2000122 (2020).

18. P. Wang, J. Liang, and L. V. Wang, "Single-shot ultrafast imaging attaining 70 trillion frames per second," Nat. Commun. **11**, 2091 (2020).
19. Z. Li, X. Jiang, S. Liu, T. Huang, J. Zheng, J. Yang, S. Li, L. Guo, X. Zhao, H. Du, T. Song, R. Yi, Y. Liu, S. Jiang, and Y. Ding, "A novel flat-response x-ray detector in the photon energy range of 0.1–4 keV," Rev. Sci. Instrum. **81**, 073504 (2010)
20. X. Liao, H. Li, and L. Carin, "Generalized alternating projection for weighted-L (2,1) minimization with applications to model-based compressive sensing," SIAM J. Imaging Sci. **7**(2), 797–823 (2014).
21. K. Dabov, A. Foi, V. Katkovnik, and K. Egiazarian, "Image denoising by sparse 3D transform-domain collaborative filtering," IEEE Trans. Image Process. **16**(8),2080-2095 (2007).
22. Y. Pu, X. Luo, L. Zhang, C. Sun, Z. Hu, G. Shen, X. Wang, Q. Tang, Z. Yuan, F. Wang, D. Yang, J. Yang, S. Jiang, Y. Ding, and J. Wang, "Chunk mixing implosion experiments using deuterated foam capsules with gold dopant," Phys. Rev. E **102**, 023204 (2020).
23. Z. Wang, A.C. Bovik, H.R. Sheikh, and E.P. Simoncelli, "Image Quality Assessment: From Error Visibility to Structural Similarity," IEEE Trans. on Image Process. **13**(4),600–612(2004).